\documentclass[onecolumn,amsmath,amssymb,superscriptaddress]{revtex4-2}
\usepackage[utf8]{inputenc}
\usepackage{epsfig,amsmath,amssymb}
\usepackage{bm}
\usepackage{graphicx}
\usepackage{graphicx}
\usepackage{dcolumn,color}

\usepackage[colorlinks, linkcolor=red,anchorcolor=green,citecolor=blue]{hyperref}
\usepackage[normalem]{ulem}

\newcommand{\Tr}{\mathop{\mathrm{Tr}}}
\newcommand\diag{\operatorname{diag}}

\newcommand{\hA}{\hat{A}}
\newcommand{\hB}{\hat{B}}

\newcommand{\bl}{\biggl}
\newcommand{\br}{\biggr}

\newcommand{\nn}{\nonumber}

\begin{document}

\title{ Generalized nonlinear Langevin equation  from  quantum nonlinear projection operator}
\author{Jin Hu}
\email{hu-j23@fzu.edu.cn}
\affiliation{Department of Physics, Fuzhou University, Fujian 350116, China}

\begin{abstract}
We systematically derive the quantum generalized  nonlinear Langevin equation using Morozov's  projection operator method. This approach extends the linear Mori-Zwanzig projection operator technique, allowing for the inclusion of nonlinear interactions among macroscopic modes. Additionally, we obtain the quantum generalized Fokker-Planck equation within the Heisenberg picture, which is consistent with Morozov's original formulation. These equations are fundamentally significant in non-equilibrium statistical physics, particularly in scenarios characterized by enhanced fluctuations, such as anomalous transport phenomena near critical points. The quantum nature of the derived generalized Langevin and Fokker-Planck equations is anticipated to provide a more detailed description than their classical equivalents.  Specifically, the noise kernel in the quantum generalized Langevin equation is multiplicative, which broadens the applicability beyond Gaussian approximations. Given specific interactions, these equations are expected to be instrumental in investigating critical transport phenomena.
  \end{abstract}

\maketitle

\section{Introduction}
  In  the phase diagram of quantum chromodynamics (QCD), the critical point (CP) is  the endpoint of the first-order phase transition line in the temperature-baryon chemical potential plane.  The properties of the critical point are among the most intriguing aspects of the QCD phase diagram, attracting considerable interest from the heavy-ion collision research community.  Since the concept of the QCD critical point was first proposed,   there has been a sustained and dedicated effort to study it theoretically. Currently, the existence of the critical point  is predicted by various QCD effective models and  suggested by lattice QCD simulations \cite{Stephanov:1998dy,Stephanov:1999zu,Fodor:2004nz,Allton:2005gk,Gavai:2008zr,deForcrand:2008zi,Hatsuda:2006ps,Braun-Munzinger:2015hba,Braun-Munzinger:2008szb,Fukushima:2013rx}. In light of these existing studies, the Beam Energy Scan (BES) program was initiated by the Relativistic Heavy Ion Collider (RHIC) at Brookhaven National Laboratory with the objective of identifying the conjectured QCD critical point.  Recent relativistic heavy-ion collision experiments at RHIC \cite{Odyniec:2013aaa}, the Facility for Antiproton and Ion Research (FAIR) \cite{STAR_NOTE_26,Senger:2022bjo}, and the Nuclotron-base Ion Collider fAcility (NICA) \cite{Kekelidze:2016hhw,Burmasov:2022clm} are dedicated to identifying the QCD critical point and have witnessed considerable experimental advancement \cite{ NA49:2007weq,STAR:2013gus, Mackowiak-Pawlowska:2014ipa,NA61SHINE:2015uhh, PHENIX:2015tkx,Luo:2015ewa,Luo:2017faz,Esha:2017dce,STAR:2017tfy,Prokhorova:2018tcl,Xu:2018vnf,Andronov:2018bln,Behera:2018wqk,STAR:2020tga,STAR:2020ddh,STAR:2021iop,STAR:2021fge,STAR:2021rls,Aparin:2022jok}.
At the critical point, very large fluctuations  occur, which could induce a critical phenomenon of divergent bulk viscosity \cite{Karsch:2007jc} as a potential signal of  the critical point's existence in experiments. The underlying mechanism driving the critical divergence of the transport coefficients is the nonlinear interactions between fluctuations:  
the transport coefficients thus can be  separated into two parts, the rapidly  relaxing part on a microscopic time scale determined by a microscopic theory, such as the Boltzmann equation \cite{DeGroot:1980dk}, and the long-time tail on a  macroscopic time scale.  Only by taking into account nonlinear interactions between macroscopic modes can the divergent behavior of the transport coefficients be fully understood \cite{Fujisaka,KAWASAKI19701}, especially when the thermal fluctuations of long-wavelength modes  are inherently large near the critical point.  This motivates the present study on the quantum generalized nonlinear Langevin equation. It should be noted that the resulting generalized nonlinear Langevin equation, due to its quantum nature, is expected to convey a richer array of physical information than its classical counterpart. The classical equation has been extensively applied to elucidate critical transport phenomena across diverse domains \cite{1977Theory}, including the QCD phase transition. The quantum framework promises to offer deeper insights into these phenomena. 

In this paper, we utilize the nonlinear projection operator method developed by
Morozov~\cite{Morozov:1981,Morozov:1992} to derive the generalized nonlinear Langevin equation in a fully quantum manner. 
Before proceeding, it is necessary to  briefly  introduce  the adopted projection formalism. The projection operator, initially proposed by Mori and Zwanzig \cite{Mori:1965,zwanzig}, has long been a powerful tool for extracting slow dynamics.  In many cases of interest,  particularly when the behavior of a system over long distances and time scales is under consideration, there is a clear separation in the time scales between  relevant slow variables and the vast number of fast microscopic degrees of freedom. The distinction between what is slow and what is fast is not that arbitrary; rather, it is determined by fundamental physical principles. For instance, the conservation laws give rise to slow variables. According to Noether's theorem, every continuous symmetry  is associated with a conserved observable that generates the symmetry. The dynamics of the local density associated with the conserved generator exhibits slow behavior, as is manifestly shown in hydrodynamics \cite{Landau:Fluid}.
A further significant source of slow variables can be found in phase transitions. As one passes into the ordered state in a phase transition, the slow variables associated with the breaking of a continuous symmetry emerge, known as Nambu–Goldstone modes \cite{Goldstone:1961eq,Nambu:1961tp}. To illustrate, the structure of the fluctuation function of the  transverse staggered fields in isotropic antiferromagnet  is qualitatively altered  by the broken symmetry, bearing resemblance to the hydrodynamic correlation functions observed in a fluid.  This indicates that the Nambu–Goldstone modes associated with the spontaneous symmetry breaking have the dynamical effect of simulating a conservation law \cite{Gene:1980dk}. 
A third example of a slow variable is the order parameter near a second-order phase transition, which gives rise to a phenomenon known as critical slowing down \cite{1977Theory}.  Furthermore, the time scale separation observed in Brownian motion \cite{Chandrasekhar:1943ws,RevModPhys.17.323}  is typically attributed to the significant difference in mass between the pollen particle and the background medium particles. Note that the formalism of the projection operator has been widely applied to study slow dynamics  in the context of condensed matter physics and modern statistical physics, see \cite{zwanzig,onuki,Gene:1980dk,balucani2003dynamical,grabert} and the reference therein for a comprehensive overview. However, Mori's original projection operator is inherently linear in the relevant variables and only applicable to systems in the linear regime \cite{Minami:2012hs}. It is important to note that the nonlinear effects are entirely encapsulated within the noise kernel, closely intertwined with the underlying microscopic dynamics. As a result, these nonlinear aspects are inherently challenging to manage.

To address this issue, the nonlinear projection operator has been proposed with significant contributions in this regard  \cite{Fujisaka,zwanzig:1975,grabert:1980,ZUBAREV1983411}.
 As will be demonstrated in the subsequent sections, this elegant approach  allows for the generation of the nonlinear coupling between the relevant macroscopic variables. The introduction of Dirac delta functions, with relevant macroscopic variables as their arguments, allows for the simultaneous derivation of both the generalized Fokker-Planck equation and the generalized Langevin equation, while maintaining exactness. The nonlinear projection operator proposed by Morozov is an extension to the quantum context,  taking into account the non-commutativity inherent in quantum operators. In this case, the traditional Dirac delta function is elevated to the status of a Dirac delta operator, in accordance with the Weyl correspondence principle \cite{Balescu,weyl:1928}, thus aligning with its classical counterpart.. 
Further technical details are provided in the main body of the text.

This paper is organized as follows. In Sec.~\ref{nonlinear},
 we present a brief review of Morozov's quantum nonlinear projection operator. In Sec.~\ref{fokker} and subsequent \ref{lagevin3}, the generalized Fokker-Planck and Langevin equations are derived using the formalism of the nonlinear projection operator. In Sec.~\ref{arkov}, we delve into the renowned Markov approximation, elucidating its foundation in the projection operator formalism.
 Finally, we give a summary and outlook in Sec.~\ref{summary}. An immediate application of the nonlinear Langevin equation to stochastic hydrodynamics is left in  Appendix.~\ref{hydro}. Throughout the script, we adopt the natural units $\hbar=k_B=c=1$ are used. The metric tensor here is given by $g^{\mu\nu}=\diag(1,-1,-1,-1)$, while $\Delta^{\mu\nu} \equiv g^{\mu\nu}-u^\mu u^\nu$ is the projection tensor orthogonal to the four-vector fluid velocity $u^\mu$.
 In addition, we employ the symmetric shorthand notations:
 \begin{eqnarray}
 X^{( \mu\nu ) } &\equiv& (X^{ \mu\nu } + X^{ \nu \mu})/2, \\
 X^{\langle \mu\nu \rangle}&\equiv&
 \bigg(\frac{\Delta^{\mu}_{\alpha} \Delta^{\nu}_{\beta} 
 	+ \Delta^{\nu}_{\alpha} \Delta^{\mu}_{\beta}}{2}
 - \frac{\Delta^{\mu\nu} \Delta_{\alpha\beta}}{3}\bigg)X^{\alpha\beta}.
 \end{eqnarray}

\section{Quantum nonlinear projection operator method}
\label{nonlinear}

In this section, we provide a brief review of Morozov's nonlinear projection operator method \cite{Morozov:1981}, which can be elegantly used  to extract 
the slow dynamics from the microscopic Hamiltonian. In quantum mechanics,  the evolution of  an operator at time $t$, $\hB(t)=e^{i\hat{H}t}\hB(0)e^{-i\hat{H}t}$, is governed by the Heisenberg equation, 
\begin{equation}
\partial_t \hB (t) = i[\hat{H},\hB (t)] \equiv iL\hB (t),
\end{equation}
where the Hamiltonian $\hat{H}$ encodes all the information of microscopic dynamics and the Liouville operator $L$ is introduced. Formally,  $\hB(t)$ can be expressed as  $\hB(t)=e^{iLt}\hB(0)$. To discuss a many-body statistical system,  it is necessary to consider a density operator, designated as $\hat{\rho}(\Gamma,t)$, which describes the phase space distribution of a given ensemble system.  The Liouville equation governing the evolution of the density operator is written as
\begin{align}
\label{liouville}
\partial_t \hat{\rho} (t)  = -iL\hat{\rho} (t),
\end{align}
where the dependence on $\Gamma$ is omitted for conciseness. As an aside, $\hat{\rho}(t)$ is sometimes referred to as  the nonequilibrium statistical operator.
Formally, $\hat{\rho}(t)$ takes the form $\hat{\rho}(t)=e^{-iLt}\hat{\rho}(0)$.
The average of $\hat{B}$ over the density operator $\hat{\rho}(t)$ is defined as
\begin{align}
\label{average}
\langle\hat{B}\rangle=\Tr(\hat{\rho}(t)\hat{B})=\Tr(\hat{\rho}\hat{B}(t)\,),
\end{align}
and the second equality indicates that  the same expectation value is given in both the Schrödinger picture and the Heisenberg picture. Equation (\ref{average}) also means that  $iL$ is an anti-self-adjoint operator in the sense of
\begin{align}
\Tr\big(\hat{B}iL\hat{C}\big)=-\Tr\big(\hat{C}iL\hat{B}\big).
\end{align}
In the remainder of this text, we choose to work in the Heisenberg picture without further  explanation, differing from the one used in \cite{Morozov:1981}. 


The physical contemplation must be initiated right from the outset, which lies in a sensible consideration of how to choose the basis variables that are projected onto. A  key point is   identifying   the slow dynamic variables. For instance, considering that conservation laws  naturally serve as a guiding principle, 
a set of coarse-grained collective variables for describing slowly varying macroscopic processes in the system are chosen as the basis vectors $\{ \hA (t, \bm{x})\} = \{ \hA_1, \hA_2, ... ,\hA_N \}$. This set includes all the local densities of the corresponding conserved quantities, which are closely related to the slow modes due to their conserved properties (other non-conserved but slow modes noted in the Introduction can also be added to the list).  Note that these properly chosen collective variables are all Hermitian. 

Mori's linear projection operator formalism  has no access to the nonlinear interactions between the slow modes because the nonlinear effects are contained in the noise kernel, rendering them intractable using this method. To address this issue,  a Dirac delta operator is introduced
\begin{align}
\hat{f}(a)=\delta (\hat{A}-a)=\frac{1}{(2\pi)^N}\int dx\exp(i\sum_{n=1}^{N}x_n (\hat{A}_n-a_n)),
\end{align}
 which is obtained according to the Weyl correspondence rule \cite{Morozov:1981,Balescu,weyl:1928} consistent with its  classical form
\begin{align}
f_{cl}(a)=\delta(A-a)=\Pi_{n=1}^{N}\delta(A_n-a_n),
\end{align}
where $N$ represents the number of basis variables. For compactness, we employ the shorthand notation $x (\hat{A}-a)$ in place of  $\sum_{n=1}^{N}x_n (\hat{A}_n-a_n)$, as needed in the subsequent discussion. Using the Dirac delta operator,  an arbitrary operator can be expressed in the following form, facilitated by the Weyl symbol,
\begin{align}
g(\hat{A})=\int da g(a)\delta(\hat{A}-a),
\end{align}
where $g(a)$ is the Weyl symbol of $g(\hat{A})$. It is important to note that $\delta(\hat{A}-a)$  can generate all nonlinear  couplings among the variables $\hat{A}_i$ effortlessly, provided $g(a)$ is a nonlinear function of $a_i$. 

In addition, we  introduce an object
\begin{align}
f(a,t)\equiv\Tr\big(\hat{\rho}(t)\hat{f}(a) \big),
\end{align}
which is the macroscopic probability density function in $a$ space. Its evolution is dictated by the generalized Fokker-Planck equation as will be shown later. 
One can also transfer to the Heisenberg  picture 
\begin{align}
\label{fat0}
f(a,t)=\Tr\big(\hat{\rho}\hat{f}(a,t) \big),
\end{align}
where  $\hat{f}(a,t)\equiv e^{iLt}\hat{f}(a)=e^{iHt}\hat{f}(a)e^{-iHt}$ is an operator in the Heisenberg picture. For a pedagogical reason,  it is more convenient to rewrite $f(a,t)$ in a classical form 

\begin{align}
f(a,t)\equiv \int d\Gamma \delta(A(\Gamma)-a)\rho(\Gamma,t)
\end{align}
where $\Gamma$ denotes the phase space variables. The phase space integral  only explores the points lying on the hypersurface constrained by $A(\Gamma)=a$. 

In the Heisenberg representation,  it is observed that for all subsequent times, $\hat{\rho}=\hat{\rho}(0)$ where $\hat{\rho}(0)$ is the initial density operator defined in the Schrödinger picture. This time independence is  particularly useful  if the system is initially in a special state. Following  \cite{Morozov:1981}, we presume that the initial state is in a state of local equilibrium $\hat{\rho}(0)=\hat{\rho}_q$ and $\hat{\rho}_q$ is typically characterized by the basis variables set, namely, $\hat{\rho}_q=\hat{\rho}_q(\hat{A}_1,\cdots \hat{A}_N)$.  As we will see below, this selection eliminates the contribution from the initial value of the noise part, i.e., the initial value that cannot be described solely by  the set of basis vectors $ \{ \hA_1, \hA_2, ... ,\hA_N \}$. Consequently, the generalized Fokker-Planck equation simplifies to a closed-form equation.   

Next, we shall delineate the construction of a quantum nonlinear projection operator. By definition, this object projects any arbitrary physical quantity  onto the relevant space expanded by the relevant set of basis vectors $ \{ \hA_1, \hA_2, ... ,\hA_N \}$.  According to Zubarev's formalism \cite{Zubarev}, the time-dependent local equilibrium density operator $\hat{\rho}_q(t)$ will always be  characterized by this relevant basis set $ \{ \hA_1, \hA_2, ... ,\hA_N \}$, with its nontrivial temporal evolution encapsulated within the conjugate fields or coefficients. Within the purview of the projection operator, the local equilibrium density operator is invariably confined to the relevant subspace. It is natural to expect that  the projection operator $P$ acts such that  $P\hat{\rho}(t)=\hat{\rho}_q(t)$,  which aligns with  Morozov's original construction (note both operators are now in the Schrödinger picture).  As will be demonstrated below, the definition Eq.(\ref{defp}) indeed satisfies this requirement.

In the Heisenberg picture, $\hat{\rho}_q$ can be parametrized  using the Weyl correspondence rule in the following form 
\begin{align}
\label{rhoq}
\hat{\rho}_q=\int da^\prime G(a^\prime)\hat{f}(a^\prime).
\end{align}

At this stage, our objective is to determine  the Weyl symbol of $\hat{\rho}_q$, $G(a^\prime)$. To achieve this, we multiply Eq.(\ref{rhoq}) by $\hat{f}(a)$ and use Eq.(\ref{fat0}),  resulting in
\begin{align}
\label{fat}
f(a)=\int da^\prime W(a,a^\prime)G(a^\prime),
\end{align}
with the definition
\begin{align}
W(a,a^\prime)\equiv\Tr(\hat{f}(a)\hat{f}(a^\prime)\,),
\end{align}
where we  use the shorthand notation $f(a)\equiv f(a,0)$.


From Eq.(\ref{fat}), we get 
\begin{align}
\label{gat}
G(a)=\int da^\prime W_{-1}(a,a^\prime)f(a^\prime),
\end{align}
where the following identity is satisfied
\begin{align}
\label{deltaw}
\int da^{\prime\prime}W(a,a^{\prime\prime})W_{-1}(a^{\prime\prime},a^\prime)=\delta(a-a^\prime).
\end{align}
Substituting Eq.(\ref{gat}) into (\ref{rhoq}), $\hat{\rho}_q$ is brought to the form
\begin{align}
\label{rhoq1}
\hat{\rho}_q=\int da da^\prime \hat{f}(a)W_{-1}(a,a^\prime)f(a^\prime).
\end{align}
To isolate the singular parts $\delta(a-a^\prime)$ in $W(a, a^\prime)$ and $W_{-1}(a, a^\prime)$, we assume
\begin{align}
\label{waa}
W(a,a^\prime)&=W(a)\big(\delta(a-a^\prime)-R(a,a^\prime)\big),\\
\label{winverse}
W_{-1}(a,a^\prime)&=W^{-1}(a^\prime)\big(\delta(a-a^\prime)+r(a,a^\prime)\big),
\end{align}
with
\begin{align}
\label{wa}
&W(a)\equiv\int da^\prime W(a,a^\prime)=\Tr(\hat{f}(a)\,),
\end{align}
where  the regular terms appearing above originate from the non-commutativity of quantum operators.
By comparing Eq.(\ref{wa}) with Eq.(\ref{waa}), we find the following relations
\begin{align}
\label{Ra}
\int da W(a)R(a,a^\prime)&=\int da^\prime R(a,a^\prime)=0,
\end{align}
and the substitution of Eqs.(\ref{waa}) and (\ref{winverse}) into Eq.(\ref{deltaw}) leads to
\begin{align}
\label{iterate1}
 r(a,a^\prime)= R(a,a^\prime)+\int da^{\prime\prime} R(a,a^{\prime\prime}) r(a^{\prime\prime},a^\prime).
\end{align}
Given $R(a,a^\prime)$, $r(a,a^\prime)$ is determined through iteration, then  Eq.(\ref{Ra}) implies 
\begin{align}
\label{ra}
\int da W(a)r(a,a^\prime)&=\int da^\prime r(a,a^\prime)=0,
\end{align}
where the second equality sign follows from the iteration of the second equality sign of Eq.(\ref{Ra}). 

In the context of the derivation presented in \cite{Morozov:1981}, Eq.(\ref{rhoq1}) plays a central role in defining the projection operator. The author utilizes this to calculate the evolution equation of $\hat{\rho}_q(t)$ (in the Schrödinger picture) and  the projection operator is naturally introduced to rewrite the resulting equation in a compact form. Following \cite{Morozov:1981},   the projection operator $P$ can be defined  as
\begin{align}
\label{defp}
P\hat{B}=\int da da^\prime \hat{f}(a)W_{-1}(a,a^\prime)\Tr(\hat{B}\hat{f}(a^\prime)\,).
\end{align}
As can be seen clearly, this definition is independent of the picture in which one works. Furthermore, one can verify that  this definition  satisfies the relation $P\hat{\rho}(t)=\hat{\rho}_q(t)$ (notice that $\hat{\rho}(0)=\hat{\rho}_q$). If $W_{-1}(a,a^\prime)\sim \delta(a-a^\prime)$ is local, 
then $P$ reduces to  a  form similar to the classical delta projection operator \cite{PhysRev.124.983}. This correspondence is not surprising: assuming the initial state is the microcanonical distribution  $f(a^\prime,0)=\delta(a^\prime-a_0)$, then in the local approximation
\begin{align}
\hat{\rho}_q=\delta(\hat{A}-a_0)/W(a_0),
\end{align}
which coincides with Eq.(12) of \cite{PhysRev.124.983}.

 It is straightforward to verify that $P$ is  idempotent, i.e., $P^2=1$.  Moreover, the projection operator satisfies
\begin{align}
\label{p1}
P\hat{f}(a)=\hat{f}(a), \\
%
\label{p3}
 \Tr( \hat{B}P\hat{C})=\Tr(\hat{C}P\hat{B}),\\
 \label{p4}
PG(\hat{A})=G(\hat{A}),
\end{align}
where the first two equalities are almost self-evident. Here $G(\hat{A})$ represents a nonlinear functional of $\hat{A}$ and a proof of the last relation is given in Eq.(\ref{ga}). Equation (\ref{p4}) highlights the nonlinearity of $P$, which extends $P\hat{A}=\hat{A}$ to $PG(\hat{A})=G(\hat{A})$, thereby incorporating the neglected nonlinear components in the projection operator.

\section{Generalized Fokker-Planck equation}
\label{fokker}

This section is dedicated to deriving the generalized Fokker-Planck equation. We begin with
\begin{align}
\label{evofa}
\frac{\partial \hat{f}(a,t)}{\partial t}=e^{iLt}iL\hat{f}(a).
\end{align}
Recall the Dyson-Duhamel  identity \cite{SCHILLING20221}
\begin{align}
e^{iLt}=e^{iLt}P+\int_0^t du e^{iLu}PiL(1-P)e^{iL(1-P)(t-u)}+(1-P)e^{iL(1-P)t},
\end{align}
which can be derived  using Laplace transformation, as demonstrated below. Initially,  the following decomposition is employed:
\begin{align}
\partial_t e^{i L t} &= e^{i L t} iL 
= e^{i L t} P iL +e^{i L t}  (1-P) iL. \label{eq:oi2}
\end{align}
Next, the Laplace transform of $\exp({i L t})$ leads us to
\begin{equation}
\int_{0}^{\infty} dt e^{-z t} e^{i L t} = \frac{1}{z - i L }.
\label{Laplace}
\end{equation}
Subsequently, a decomposition of  Eq.~(\ref{Laplace}) into
\begin{align}
\frac{1}{z - i L } &= \frac{1}{z - i L }(z - (1-P) iL)\frac{1}{z - (1-P) iL} \notag \\
&= \frac{1}{z - i L }(z - iL + P i L)\frac{1}{z - (1-P) iL} \notag \\
&= \frac{1}{z -(1-P) iL } + \frac{1}{z - i L} P iL \frac{1}{z -(1-P) iL }
\end{align}
is utilized. After performing the inverse Laplace transform,  we arrive at a crucial identity
\begin{align}
\label{eilt}
e^{i L t}=e^{(1-P) i L t}+\int_{0}^{t}ds e^{i L (t-s)} P iL e^{(1-P) i L s},
\end{align}
which follows from the derivation in \cite{Minami:2012hs}.   Noticing that 
\begin{align}
(1-P)e^{iL(1-P)t}=e^{(1-P)iLt}(1-P),
\end{align}
 multiplying
Eq.(\ref{eilt}) by $(1-P)$ from the right
yields the desired Dyson-Duhamel identity. This completes the proof.

It is time to apply Dyson-Duhamel  identity to the right-hand side of Eq.(\ref{evofa})
\begin{align}
\label{fatt}
\frac{\partial \hat{f}(a,t)}{\partial t}=e^{iLt}PiL\hat{f}(a)+\int_0^t du e^{iLu}PiL(1-P)e^{iL(1-P)(t-u)}iL\hat{f}(a)+(1-P)e^{iL(1-P)t}iL\hat{f}(a).
\end{align}

To proceed, we introduce a useful definition
\begin{align}
iL\hat{f}(a)\equiv-\frac{\partial \hat{J}_i(a)}{\partial a_i},
\end{align}
with
\begin{align}
\hat{J}(a)\equiv\frac{1}{(2\pi)^N}\int dx e^{ix(\hat{A}-a)}\int_0^1 d\tau e^{-i\tau x\hat{A}}iL\hat{A}e^{i\tau x\hat{A}}.
\end{align}
where the Einstein summation convention is implied and the summation symbol will be omitted  hereafter for compactness, when nothing confusing occurs. In the derivation of this equation, we utilize an important identity known as the Kubo identity \cite{kubo}
\begin{align}
[e^{\hat{B}},\hat{H}]=e^{\hat{B}}\int_0^1 d\tau e^{-\tau\hat{B}}[\hat{B},\hat{H}]e^{\tau\hat{B}}.
\end{align}

The first term in Eq.(\ref{fatt}) is
\begin{align}
e^{iLt}PiL\hat{f}(a)&= -\int da^{\prime\prime} da^\prime W_{-1}(a^{\prime\prime},a^\prime)\Tr(\frac{\partial \hat{J}_i(a)}{\partial a_i}\hat{f}(a^\prime)\,)\hat{f}(a^{\prime\prime},t)\nn\\
&= -\frac{\partial}{\partial a_i}\int da^\prime v_i(a,a^{\prime})\hat{f}(a^{\prime},t),
\end{align}
with the nonlocal streaming velocity
\begin{align}
v_i(a,a^\prime)&\equiv\int da^{\prime\prime}W_{-1}(a^\prime,a^{\prime\prime})\Tr(\hat{J}_i(a)\hat{f}(a^{\prime\prime})\,).
\end{align}

Then, by defining
\begin{align}
\hat{X}_i(a)=(1-P)\hat{J}_i(a),
\end{align}
we can express the noise term as
\begin{align}
(1-P)e^{iL(1-P)t}iL\hat{f}(a)=-\frac{\partial}{\partial a_i}\hat{X}_i(a,t).
\end{align}
The remaining diffusion term is 
\begin{align}
&PiL(1-P)e^{iL(1-P)(t-u)}iL\hat{f}(a)=PiLe^{i(1-P)L(t-u)}(1-P)iL\hat{f}(a)\nn\\
&=\frac{\partial}{\partial a_i} \int da^\prime K_{ij}(a,a^\prime,t-u)\frac{\partial}{\partial a^\prime_j}\int da^{\prime\prime}\hat{f}(a^{\prime\prime}) W_{-1}(a^{\prime\prime},a^\prime).
\end{align}
where $\hat{X}(a,t)\equiv e^{(1-P)iLt}\hat{X}(a)$. The diffusion kernel  $K$  is intricately to the noise function $X$ through a generalized fluctuation-dissipation theorem $K_{ij}(a,a^\prime,t)\equiv\Tr(\hat{X}_i(a,t)\hat{X}_j(a^\prime)\,)$, which holds profound theoretical significance in the realm of statistical physics.

Putting them together, we arrive at  the quantum Fokker-Planck equation in operator form 
\begin{align}
\label{fatt1}
\frac{\partial \hat{f}(a,t)}{\partial t}=&-\frac{\partial}{\partial a_i}\int da^\prime v_i(a,a^{\prime})\hat{f}(a^{\prime},t)+\int_0^t du \frac{\partial}{\partial a_i} \int da^\prime K_{ij}(a,a^\prime,t-u)\frac{\partial}{\partial a^\prime_j}\int da^{\prime\prime}\hat{f}(a^{\prime\prime},u) W_{-1}(a^{\prime\prime},a^\prime)\nn\\
&-\frac{\partial}{\partial a_i}\hat{X}_i(a,t),
\end{align}
which is consistent with the Fokker-Planck equation obtained in a fully classical way \cite{Fujisaka}. The fluctuating force $\frac{\partial}{\partial a_i}\hat{X}_i(a,t)$ is uncorrelated with an arbitrary function $G(\hat{A})$ of $\hat{A}$, 
\begin{align}
\label{Ga}
\Tr(G(\hat{A})\hat{X}(a,t))=0.
\end{align}

Next we calculate the average of Eq.(\ref{fatt1}) with respect to $\hat{\rho}$,
\begin{align}
\label{fatt5}
\frac{\partial f(a,t)}{\partial t}=&-\frac{\partial}{\partial a_i}\int da^\prime v_i(a,a^{\prime})f(a^{\prime},t)+\int_0^t du \frac{\partial}{\partial a_i} \int da^\prime K_{ij}(a,a^\prime,t-u)\frac{\partial}{\partial a^\prime_j}\int da^{\prime\prime}f(a^{\prime\prime},u) W_{-1}(a^{\prime\prime},a^\prime),
\end{align}
where the expectation value of the noise term vanishes because
\begin{align}
\Tr[\hat{\rho}(1-P)e^{iL(1-P)t}iL\hat{f}(a)]=&\Tr[\hat{\rho}_q(1-P)e^{iL(1-P)t}iL\hat{f}(a)]=\Tr[((1-P)\hat{\rho}_q) e^{iL(1-P)t}iL\hat{f}(a)]=0.
\end{align}
Here we adopt the assumption $\hat{\rho}=\hat{\rho}_q$ within the Heisenberg picture. 
Equation (\ref{fatt5}) is in agreement with the Fokker-Planck equation presented in \cite{Morozov:1981}, which was  derived in a different manner :  solving $\hat{\rho}(t)$ directly from the retarded Liouville equation and subsequently constructing $f(a,t)$ based on  this solution. The identical results obtained from two distinct quantum pictures confirm their consistency. Furthermore, a similar Fokker-Planck equation has been derived in a classical context, as demonstrated in \cite{PhysRev.124.983}. While numerous other studies have also explored this equation, they are not all cited here due to space constraints. 
In this script, we refer to the equation as the Fokker-Planck equation in functional form. 


Let us pause for a moment  to derive a useful relation. The trace over $\hat{f}(a,t)$ is expected to remain invariant over time, as expressed by the equation
\begin{align}
\frac{\partial W(a)}{\partial t} =0 \quad \text{with}\quad W(a)\equiv \Tr(\hat{f}(a,t)\,),
\end{align}
owing to the cyclical symmetry of the trace $W(a)=\Tr(\hat{f}(a)\,)$.  It is crucial to distinguish between $f(a,t)=\Tr(\hat{\rho}\hat{f}(a,t))$ and $W(a)=\Tr(\hat{f}(a,t)\,)$ to avoid confusion. Subsequently, taking the trace over the entire Fokker-Planck equation (\ref{fatt1}), we obtain
\begin{align}
\label{fatt2}
\frac{\partial W(a)}{\partial t} =&-\frac{\partial}{\partial a_i}\int da^\prime v_i(a,a^{\prime})W(a^{\prime})+\int_0^t du \frac{\partial}{\partial a_i} \int da^\prime K_{ij}(a,a^\prime,t-u)\frac{\partial}{\partial a^\prime_j}\int da^{\prime\prime}W(a^{\prime\prime}) W_{-1}(a^{\prime\prime},a^\prime)\nn\\
&-\frac{\partial}{\partial a_i}\Tr(\hat{X}_i(a,t)\,).
\end{align}
Our remaining task is to  determine whether the right-hand side (RHS) is naturally zero. If not,  then Eq.(\ref{fatt2}) imposes additional constraints.

The first term on the RHS corresponds to the divergence condition in \cite{Kim:1991},
\begin{align}
&\frac{\partial}{\partial a_i}\int da^\prime v_i(a,a^{\prime})W(a^{\prime})=\frac{\partial}{\partial a_i}\int da^\prime\int da^{\prime\prime}\Tr(\hat{J}_i(a)\hat{f}(a^{\prime\prime})\,)W_{-1}(a^\prime,a^{\prime\prime})W(a^{\prime})\nn\\
=&\frac{\partial}{\partial a_i}\int da^\prime\Tr(\hat{J}_i(a)\hat{f}(a^{\prime})\,)=\frac{\partial}{\partial a_i}\Tr(\hat{J}_i(a)\,)=-\Tr(iL\hat{f}(a)\,)=i\Tr([\hat{f}(a),\hat{H}])=0.
\end{align}
According to the above calculation,  this  term  vanishes unconditionally. Therefore, it should not  be regarded as a  constraint condition. A similar conclusion is reached within the framework of classical statistical physics, as detailed in \cite{Gene:1980dk}. The second term  also vanishes because it is a constant
\begin{align}
\int da^{\prime\prime}W(a^{\prime\prime}) W_{-1}(a^{\prime\prime},a^\prime)=1+W^{-1}(a^\prime)\int da^{\prime\prime}W(a^{\prime\prime}) r(a^{\prime\prime},a^\prime)=1,
\end{align}
 where the condition in Eq.(\ref{ra}) is used. As for the last noise term, we provide a  proof of its vanishing  in Appendix.\ref{vanishnoise}.
 
In conclusion, $\frac{\partial W(a)}{\partial t}=0$ holds naturally without imposing additional constraints. Notably,  such constraints are crucial in constructing stochastic hydrodynamic equations in a classical context \cite{Kim:1991}. 
The authors there  treat  $\frac{\partial W(a)}{\partial t}=0$  as a useful  constraint condition, from which  the stochastic hydrodynamic equations with multiplicative noises  can be worked out. However, based on our derivation presented above,  there are no constraint conditions originating from Eq.(\ref{fatt2}) because it holds unconditionally. Note as an aside, this conclusion is also  confirmed by one of the authors of \cite{Kim:1991} in a later textbook \cite{Gene:1980dk} using a classical treatment. 
Therefore, we must strictly adhere to Eq.(\ref{lagevin}), and  explore possible extensions to include multiplicative noises within hydrodynamic framework. 


\section{Generalized Langevin equation}
\label{lagevin3}
In this section, we derive the generalized nonlinear Langevin equation from the generalized Fokker-Planck equation (\ref{fatt1}). By  multiplying Eq.(\ref{fatt1}) by $a$ and then integrating over $a$,  we  find
\begin{align}
\label{langevin0}
\frac{\partial \hat{A}_i(t)}{\partial t}=&-\int da a_i \frac{\partial}{\partial a_j}\int da^\prime v_j(a,a^{\prime})\hat{f}(a^{\prime},t)+\int da a_i\int_0^t du \frac{\partial}{\partial a_l} \int da^\prime K_{lj}(a,a^\prime,t-u)\nn\\
&\times\frac{\partial}{\partial a^\prime_j}\int da^{\prime\prime}\hat{f}(a^{\prime\prime},u) W_{-1}(a^{\prime\prime},a^\prime)-\int da a_i\frac{\partial}{\partial a_j}\hat{X}_j(a,t)\nn\\
=&\int da \int da^\prime v_i(a,a^{\prime})\hat{f}(a^{\prime},t)+\int_0^t du   \int da^\prime (\frac{\partial}{\partial a^\prime_j}K_{i j}(a^\prime,t-u)\,) \int da^{\prime\prime}\hat{f}(a^{\prime\prime},u) W_{-1}(a^{\prime\prime},a^\prime)+\hat{R}_i(t)
\end{align}
where $\hat{R}_k(t)\equiv e^{(1-P)iLt}(1-P)iL\hat{A}_k$, and  we recall that
\begin{align}
\hat{X}_i(a)=(1-P)\hat{J}_i(a)=\frac{1}{(2\pi)^N}(1-P)\int dx\int_0^1 d\tau e^{ix(\hat{A}-a)} e^{-i\tau x\hat{A}}(iL\hat{A}_i)e^{i\tau x\hat{A}}.
\end{align}
Here we also define
\begin{align}
K_{ij}(a^\prime,t)&\equiv \int daK_{ij}(a,a^\prime,t)=\int da\Tr(\hat{X}_i(a,t)\hat{X}_j(a^\prime)\,)\nn\\
&=\frac{1}{(2\pi)^{2N}}\int da\int dxdx^\prime\int d\tau d\tau^\prime\Tr(e^{i(1-P)Lt}(1-P)e^{ix(\hat{A}-a)} e^{-i\tau x\hat{A}}(iL\hat{A}_i)e^{i\tau x\hat{A}}\,\,\nn\\
&\times(1-P)e^{ix^\prime(\hat{A}-a^\prime)} e^{-i\tau^\prime x^\prime\hat{A}}(iL\hat{A}_j)e^{i\tau^\prime x^\prime\hat{A}}\,)\nn\\
&=\frac{1}{(2\pi)^{N}}\int dx^\prime \int_0^1d\tau^\prime\Tr\big( (e^{i(1-P)Lt}(1-P)iL\hat{A}_i)\,\,(1-P)e^{ix^\prime(\hat{A}-a^\prime)} e^{-i\tau^\prime x^\prime\hat{A}}(iL\hat{A}_j)e^{i\tau^\prime x^\prime\hat{A}}\,\big)\nn\\
&=\Tr(\hat{R}_i(t)\hat{X}_j(a^\prime)).
\end{align}
Due to Eq.(\ref{Ga}), the Langevin fluctuating force  is also statistically independent of the relevant variables, namely,
\begin{align}
\label{Ga1}
\Tr(G(\hat{A})\hat{R}_i(a,t))=0.
\end{align}
Equation (\ref{langevin0}) represents the complete quantum generalized nonlinear Langevin equation without any approximations. The first term on the RHS  is the drift term characterized by a nonlocal streaming velocity $v(a,a^\prime)$. This nonlocality arises from the non-commutativity of quantum operators, a quantum effect absent in the classical Langevin equation. The second one is a diffusion term with non-Markov memory effects. Examining the specific form of the diffusion  kernel $K_{ij}(a,t)$ reveals the involvement of quantum non-commutativity, hindering  direct comparison with its classical counterpart. The third one represents the fast motion of noise, with its relation to the diffusion kernel described by the fluctuation-dissipation theorem. 
 To accurately account for the physical effects  induced by thermal fluctuations within the fluid, the  stochastic noise term is essential \cite{POMEAU197563}.

For comparison with existing classical results, we must consider the local or classical approximation
\begin{align}
&v(a,a^\prime)=v(a)\delta(a-a^\prime),\quad W_{-1}(a^{\prime\prime},a^\prime)=W^{-1}(a^\prime)\delta(a^\prime-a^{\prime\prime}),
\end{align}
where we define the local averaged velocity as
\begin{align}
\label{va}
&v(a)=\Tr(\hat{f}(a)iL\hat{A})W^{-1}(a).
\end{align}
Thus, the nonlinear Langevin equation,  the main result of our script, becomes
\begin{align}
\label{lagevin0}
\frac{\partial \hat{A}_i(t)}{\partial t}
=&v_i(\hat{A},t)+\int_0^t du   \int da^\prime (\frac{\partial}{\partial a^\prime_j}K_{i j}(a^\prime,t-u)\,) W^{-1}(a^\prime)\hat{f}(a^{\prime},u)+\hat{R}_i(t),
\end{align}
which matches the result of \cite{Fujisaka}. Furthermore, introducing the definition $\tilde{K}_{ij}(a,t)\equiv K_{ij}(a,t)/W(a)$ transforms the equation into
\begin{align}
\label{lagevin}
\frac{\partial \hat{A}_i(t)}{\partial t}=&v_i(\hat{A},t)+\int_0^t du   \int da \frac{\partial}{\partial a_j}(\tilde{K}_{i j}(a,t-u)W(a)\,) W^{-1}(a)\hat{f}(a,u)+\hat{R}_i(t)\nn\\
=&v_i(\hat{A},t)+\int_0^t du   \int da (\frac{\partial}{\partial a_j}\tilde{K}_{i j}(a,t-u)\,) \hat{f}(a,u)+ \int_0^t du\int da \tilde{K}_{i j}(a,t-u)F_j(a)\hat{f}(a,u)+\hat{R}_i(t),
\end{align}
in good agreement with the form presented in \cite{PhysRev.124.983}. 
Here the conjugate variable of $a_k$ is defined, in a conventional way, as the derivative of $\ln W(a)$ with respect  to $a_k$
\begin{align}
\label{fka}
F_k(a)\equiv \frac{\partial}{\partial a_k}\ln W(a).
\end{align}
Since $W(a)$ provides a complete thermodynamic description of a system in thermal equilibrium,  specifying the values - “$\hat{A}=a$", it can be viewed  as a partition function of the microcanonical ensemble.  More intuitively,  Eq.(\ref{va}) can be rewritten as  
\begin{align}
\label{va1}
&v(a)=\Tr(\hat{\rho}(a)iL\hat{A}),
\end{align}
with the definition $\hat{\rho}(a)\equiv \hat{f}(a)W^{-1}(a)$. Evidently, $\Tr\hat{\rho}(a)=1$ and $\Tr(\hat{\rho}(a)\hat{A}_i\,)=a_i$,  indicating that $\hat{\rho}(a)$ acts like  a density operator in a microcanonical ensemble with fixed values of the coarse-grained variables. 
By definition, $F_k(a)$ can be interpreted as the thermodynamic force. Then the third term in the last line of  Eq.(\ref{lagevin})  can be viewed as a generalized Onsager relation in a convolution form with memory effects. 

The second term on the RHS of Eq.(\ref{lagevin}) is crucial,  demonstrating that
the diffusion kernel $\tilde{K}$ exhibits, in general, nontrivial dependence on the relevant variables. Nonetheless, this term diminishes under certain commonly applied simplifications. In this place, we aim to show a simplified but helpful form of the generalized Langevin equation, where the second term vanishes. This is achieved by substituting the diffusion kernel with its average value
\begin{align}
\tilde{K}_{i j}(a,t-u)\rightarrow \int da W(a)\tilde{K}_{i j}(a,t-u)=\int da K_{i j}(a,t-u)=\Tr(\hat{R}_i(t-u)\hat{R}_j),
\end{align}
where $W(a)$ is considered as an equilibrium distribution function for  $a$, up to a normalization constant
\begin{align}
\int da W(a)=\int da \Tr(\hat{f}(a)\,)=\Tr(1)=\text{const}.
\end{align}
This constant can be incorporated into the definitions of other coefficients, thus eliminating the need for concern over its impact.
Such a replacement is also implemented in \cite{Fujisaka} to seek a  simpler equation.  Upon completing the replacement, we cast Eq.(\ref{lagevin}) into
\begin{align}
\label{lagevin1}
\frac{\partial \hat{A}_i(t)}{\partial t}
=v_i(\hat{A},t)+ \int_0^t du \Tr(\hat{R}_i(t-u)\hat{R}_j)F_j(\hat{A}(u))+\hat{R}_i(t),
\end{align}
which recovers our familiar non-Markovian nonlinear Langevin equation. 

As seen above, the local approximation is crucial for simplification operations. This validity is assured for classical systems and may also extend to certain quantum systems as well. In physical systems of interest, such as relativistic heavy ion collisions or the early universe, the environment is anticipated to be extremely hot, leading to the suppression of quantum effects by the temperature scale. 
 Furthermore, if the system's non-equilibrium state is characterized by a slow change in the distribution function $f(a, t)$ and the corresponding density of states $W(a)$ changes slowly with respect to the variation in the distance $|a-a^\prime|$,  in comparison to the variation of $r(a, a^\prime)$, then, in the first approximation, these functions can be considered constant when calculating the integral term in the Fokker-Planck and Langevin equations. In this case, under these conditions, the contribution from the regular parts is rendered negligible due to Eq.(\ref{ra}).

\section{Markov approximation}
\label{arkov}

The Markov approximation is often necessary to analyze the late-time dynamic evolution of a system. Memory effects or initial correlations are typically eliminated after multiple collisions between particles. When discussing late-time dynamics, the Boltzmann equation or hydrodynamics are commonly utilized frameworks, where memory effects are effectively removed. For example, in casting the generalized Langevin equation into the framework of fluctuating hydrodynamics, the Markov approximation is essential for eliminating delay effects in the integral term. This section provides a detailed explanation of the Markov approximation in the context of  the projection operator.

Given that  the operator of projection $1-P$ in the diffusion kernel $K(a,t)$ or $K(a,a^\prime,t)$ precludes  the slow evolution associated with the relevant variables,  the effect of time delays  can be neglected in the evolution equations  when a clear separation of time scales exists.  This implies that the time scales required for a significant alteration of $\hat{A}(t)$ are distinctly larger than the typical time scales involved in the Langevin fluctuating force $\hat{R}_i(t)$. 

Note that Eq.(\ref{lagevin1}) can be rewritten as 
\begin{align}
\label{lagevins}
\frac{\partial \hat{A}_i(t)}{\partial t}
=\,v_i(\hat{A},t)+ \int_0^t du \Tr(\hat{R}_i(u)\hat{R}_j)F_j(\hat{A}(t-u))+\hat{R}_i(t),
\end{align}
where we perform a variable transformation. Then we must distinguish two typical timescales $t$ and $u$ involved in the diffusion term: $t$ characterizes the slow evolution of $\hat{A}(t)$, while $u$ is related to the fast motion of noise $\hat{R}_i(u)$. With a hierarchy $t\gg u$ signifying the clear scale separation, 
the  Langevin equation turns into a simplified form
\begin{align}
\label{lagevin2}
\frac{\partial \hat{A}_i(t)}{\partial t}
=\,v_i(\hat{A}(t))+ \gamma_{i j}F_j(\hat{A}(t))+\hat{R}_i(t),
\end{align}
where $\gamma$ is the bare kinetic coefficient defined as
\begin{align}
\label{flucdiss}
\gamma_{ij}\equiv \int_0^\infty du \Tr(\hat{R}_i(u)\hat{R}_j).
\end{align}
The first two terms on the RHS of Eq.(\ref{lagevin2}) typically contain nonlinear coupling of $\hat{A}$, while $\hat{R}_i(t)$ represents  a fluctuating force
satisfying
\begin{align}
\Tr(\hat{R}_i(t)\hat{R}_j(t^\prime)\,)=2\gamma_{ij}\delta(t-t^\prime).
\end{align}
As demonstrated in \cite{Fujisaka}, the nonlinear couplings in the drift term lead to the renormalization of the bare kinetic coefficient,  giving rise to their critical divergence (the nonlinearity in the diffusion term can also renormalize the bare kinetic coefficient, but the contribution would  typically be suppressed in the long-wavelength limit). Therefore, Eq.(\ref{lagevin2}) has been extensively applied to  investigate the critical transport phenomena associated with various phase transition systems.

So far, we have successfully reproduced the conventional nonlinear Langevin equation, which has found applications across diverse domains.  To maintain transparency throughout the reduction process, we have introduced approximations incrementally, clarifying which elements are preserved and which are excluded. This method is particularly advantageous for assessing the selected approximations and for reintegrating pertinent  physical insights.


\section{Summary and Outlook} \label{summary}
In this work, we present a systematic derivation of the quantum generalized nonlinear Langevin equation employing the quantum nonlinear projection operator method. 
Morozov's nonlinear projection operator, as an extension of the well-established linear Mori-Zwanzig projection operator, enables the consideration of nonlinear interactions among macroscopic modes.  This approach yields the quantum Fokker-Planck and Langevin equations, which are pivotal in investigating anomalous transport phenomena in the vicinity of the critical point. Notably, these two equations are derived in a fully quantum manner and are  expected to carry more physical information  than their classical counterparts. 
Furthermore, we elaborate on how to introduce  physical approximations step by step to reduce the obtained nonlinear Langevin equation to its conventional form. 
 
 
Some possible extensions could be made in the future. The first one is to derive the  fluctuating hydrodynamics with the multiplicative noises. Though the stochastic hydrodynamic equations with addictive noises can be effortlessly reproduced by the present formalism, they don't exhibit any significant difference. In this case, the kinetic coefficients are independent of the fluctuating fields. 
If the perturbation is not too large to change the intrinsic properties, the  field-independent kinetic coefficients suffice to characterize  the system with given thermodynamic states $(e_0,p_0\cdots)$,  in line with the logic hidden in  linear response theory. Conversely, if the perturbation is strong enough to fundamentally alter the  intrinsic properties and  the thermodynamic state of the system (e.g., the critical behavior near the critical point), the stochastic hydrodynamics with multiplicative noises and field-dependent kinetic coefficients is essentially needed. 
 With the foundational framework in place, a comprehensive dynamic renormalization group analysis can be readily pursued, which is a well-established  technique for studying critical phenomena in statistical physics.  We expect to see the application in the study of the critical behavior  in the vicinity of the QCD critical point. Furthermore, our formalism is inherently consistent with a first-principles calculation based on quantum field theory, offering a distinct advantage over the classical nonlinear Langevin equation. Inspired by the work detailed in \cite{Koide:2004yn} using the quantum linear Langevin equation, we can move forward  to include nonlinear effects within the framework and investigate the critical  transport phenomena associated with QCD phase transition. 
 Last but not least, the obtained  generalized Fokker-Planck and Langevin equations  can be helpful in condensed matter physics or other low-temperature systems, where quantum effects are non-negligible and of significant research interest. In those cases, the local approximation or the classical  approximation breaks down. Thus it would be extremely useful  to apply the quantum generalized  Fokker-Planck and Langevin equations to study the relevant transport phenomena.

\section{Acknowledgments} 
J.H. is grateful to Yuki Minami and Giorgio Torrieri for helpful correspondence and  Shuzhe Shi  for  helpful discussions. 

\begin{appendix}
	
\section{The proof of useful identities}
\label{projection}

In this appendix, we prove some useful identities associated with the nonlinear projection operator. The first one is $P\hat{A}=\hat{A}$. We have
\begin{align}
P\hat{A}&=\int da da^\prime \hat{f}(a)W_{-1}(a,a^\prime)\Tr(\hat{A}\hat{f}(a^\prime)\,)\nn\\
&=\int da da^\prime a^\prime\hat{f}(a)W_{-1}(a,a^\prime) W(a^\prime)\nn\\
&=\int da da^\prime a^\prime\hat{f}(a)(\delta(a-a^\prime)+r(a,a^\prime)\,)\nn\\
&=\hat{A}+\int da\hat{f}(a)\int da^\prime a^\prime r(a,a^\prime),
\end{align}
and
\begin{align}
\int da^\prime a^\prime W(a,a^\prime)=\int da^\prime a^\prime\Tr(\hat{f}(a)\hat{f}(a^\prime)\,)=\Tr(\hat{f}(a)\hat{A}\,)=aW(a),
\end{align}
Notice that the above equation can be expressed in a different way
\begin{align}
\int da^\prime a^\prime W(a,a^\prime)=\int da^\prime a^\prime W(a)(\delta(a-a^\prime)+R(a,a^\prime)\,)=aW(a)\rightarrow \int da^\prime a^\prime R(a,a^\prime)=0.
\end{align}
According to Eq.(\ref{iterate1}),
\begin{align}
\label{iterate2}
a^\prime r(a,a^\prime)=a^\prime R(a,a^\prime)+\int da^{\prime\prime} R(a,a^{\prime\prime})a^\prime r(a^{\prime\prime},a^\prime).
\end{align}
By iterating Eq.(\ref{iterate2}), we reach 
\begin{align}
\int da^\prime a^\prime r(a,a^\prime)=0,
\end{align}
which follows from the same reason for obtaining the second identity in Eq.(\ref{ra}).
Finally, this completes the proof of $P\hat{A}=\hat{A}$. 

The above relation can be also derived in a simpler way, noticing that
\begin{align}
P\hat{f}(a)=\hat{f}(a),
\end{align}
then an integral over $a$ constructed as
\begin{align}
\int daG(a)P\hat{f}(a)=\int da G(a)\hat{f}(a).
\end{align}
Because $P$ only acts upon the operator, it can be safely factorized out 
\begin{align}
\int daG(a)P\hat{f}(a)=P\int da G(a)\hat{f}(a),
\end{align}
this eventually gives us
\begin{align}
\label{ga}
PG(\hat{A})=G(\hat{A}).
\end{align}
Here $G(\hat{A})$ is a nonlinear functional of $\hat{A}$.

\section{The vanishing of noise contribution}
\label{vanishnoise}

The  trace  of the noise term is
\begin{align}
\Tr[(1-P)e^{iL(1-P)t}iL\hat{f}(a)]=&\Tr[e^{iL(1-P)t}iL\hat{f}(a)]-\Tr[Pe^{iL(1-P)t}iL\hat{f}(a)]\nn\\
&=\Tr[\hat{Y}(a,t)]-\Tr[\int da^{\prime\prime} da^\prime \hat{f}(a^{\prime\prime})W_{-1}(a^{\prime\prime},a^\prime)\Tr(\hat{Y}(a,t)\hat{f}(a^\prime)\,)]\nn\\
&=\Tr[\hat{Y}(a,t)]-\int da^{\prime\prime} da^\prime W_{-1}(a^{\prime\prime},a^\prime)\Tr\big(\hat{f}(a^{\prime\prime})\big)\Tr\big(\hat{Y}(a,t)\hat{f}(a^\prime)\big)\nn\\
&=\Tr[\hat{Y}(a,t)]-\int da^{\prime\prime} da^\prime W_{-1}(a^{\prime\prime},a^\prime)W(a^{\prime\prime})\Tr\big(\hat{Y}(a,t)\hat{f}(a^\prime)\big)\nn\\
&=\Tr[\hat{Y}(a,t)]-\Tr[\hat{Y}(a,t)]=0,
\end{align}
where the shorthand notation  $\hat{Y}(a,t)\equiv e^{iL(1-P)t}iL\hat{f}(a)$ is used.

\section{The derivation of stochastic hydrodynamics}
\label{hydro}

\subsection{Relativistic hydrodynamics}

The evolution of a relativistic  fluid is governed by the continuity equations
\begin{align}
\label{Tcon}
&\partial_\mu T^{\mu\nu} =0 \, ,
\\
\label{Ncon}
&\partial_\mu N^{\mu} = 0 \, ,
\end{align}
where $T^{\mu\nu}, N^{\mu}$ are the energy-momentum tensor and conserved current respectively. In this subsection, the hat symbol of the operators is temporarily  omitted.

By utilizing symmetry analysis and the second law of thermodynamics,  $T^{\mu\nu}$ and $N^{\mu}$ can be conveniently  expressed  in Landau frame
\begin{align}
&
T^{\mu\nu} = e u^\mu u^\nu - p \Delta^{\mu\nu} + \tau^{\mu\nu}
\, ,
\\
&
N^{\mu} = n u^\mu + j^{\mu}
\, ,
\end{align}
with $e$, $n$, and $p$ being the energy density, the charge density, and the pressure, respectively. The metric tensor  is given by $g^{\mu\nu}=\diag(1,-1,-1,-1)$, while $\Delta^{\mu\nu} \equiv g^{\mu\nu}-u^\mu u^\nu$ serves as the projection tensor orthogonal to the four-vector fluid velocity $u^\mu$. Also, the dissipative terms $\tau^{\mu\nu}$ and $j^\mu$  are given by
\begin{eqnarray}
&&
\label{shear}
\tau^{\mu\nu}=-2\eta\nabla^{\langle\mu}u^{\nu\rangle}-\zeta\theta \Delta^{\mu\nu},
\\
&&
\label{heat}
j^\mu=-\frac{\sigma T(e+p)}{n}\nabla^\mu\frac{\mu}{T},
\end{eqnarray}
where $\mu$ is the chemical potential conjugate to the conserved charge density $n$ and $\nabla_\mu\equiv \Delta_{\mu\nu}\partial^\nu$ is the spatial derivative. A shorthand notation for the thermodynamic force $\theta=\nabla\cdot u$ denotes the expansion rate. Three transport coefficients $\eta$, $\zeta$, and $\sigma$ are  the bare shear viscosity,  bare bulk viscosity, and  bare thermal conductivity, respectively.

In the linear regime,  hydrodynamic modes can  be analyzed using linearized equations. A  linear mode analysis  is conducted on top of the background of  thermal equilibrium.   In a relativistic fluid, the quiescent equilibrium system is  perturbed according to
\begin{align}
\label{perturb}
e (t,\bm{x}) &= e_0 + \delta e (t,\bm{x}), \quad
p (t,\bm{x}) = p_0 + \delta p (t,\bm{x}),
\nn
\\
n (t,\bm{x}) &= n_0 + \delta n (t,\bm{x}), \quad u^\mu (t,\bm{x}) = (1,0) + (0,\delta v^i (t,\bm{x})),\nn\\
T(t,\bm{x})&=T_0+\delta T(t,\bm{x}),\quad \frac{\mu}{T}=\frac{\mu_0}{T_0}+\delta(\frac{\mu}{T}).
\end{align}
Given $u\cdot u=1$, $(\delta v)^2\ll 1$, terms of $O(\delta v^2)$ are subsequently neglected.
Then we have
\begin{eqnarray}
\frac{\partial \delta n}{\partial t}=&&-n_0 \nabla \cdot \delta v 
-\frac{\sigma T_0(e_0+p_0)}{n_0}  \nabla^2 \delta 
\bl( \frac{\mu}{T}\br),
\label{eq: ln}\\
\frac{\partial \delta e}{\partial t}=&& - (e_0+p_0)\nabla \cdot \delta v , \label{eq: le}\\
\frac{\partial \pi_i}{\partial t}=&& -\nabla_i\delta p
-(\zeta+\frac{1}{3}\eta)\nabla_i(\nabla\cdot\delta v ) -\eta\nabla^{2}\delta v_i .
\end{eqnarray}
where $\pi^i\equiv \delta T^{0i}$.

\subsection{Relativistic fluctuating hydrodynamics}
\label{fluc}
The nonlinear interactions between these hydrodynamic modes are absent in conventional hydrodynamics in the linear regime \cite{Minami:2012hs}.  More importantly, without stochastic noises, conventional hydrodynamics ignores the inherent hydrodynamic excitations triggered by thermal  fluctuations. The effects of nonlinear fluctuations in relevant variables can be appropriately addressed within the framework of the Langevin equation. In this subsection, we derive the relativistic fluctuating hydrodynamics from the Langevin equation step by step. 

Inspired by relativistic  hydrodynamic equations, we consider the relevant set of slow variables as $\{ \hA (t, \bm{x})\} = \{ \delta e(t,\bm{x}),\delta n(t,\bm{x}),\pi_i(t,\bm{x}) \}$.  First, we need to figure out the  streaming term
\begin{align}
\label{drift}
v(a)=\Tr(\hat{\rho}(a)iL\hat{A}(t,\bm{x}))=\Tr(\hat{\rho}(a)\partial_t\hat{A}(t,\bm{x})).
\end{align}
Examining the continuity equation,
\begin{align}
\partial_t\hat{A}(t,\bm{x})=\nabla\cdot \hat{J}_A,
\end{align}
the derivative of $\hat{A}$ with respect to 
$t$ yields the corresponding reversible currents, with dissipative currents averaging to zero over a thermodynamic ensemble, as depicted on the right-hand side of Eq.(\ref{drift}). Therefore, we can directly read the streaming velocities,
\begin{align}
v_n=-\nabla\cdot(\hat{n}\delta \hat{v}), \quad v_e=-\nabla\cdot\hat{\pi},\quad v_{\pi}=-\nabla_i \delta \hat{p},
\end{align}
where the background is set to be at rest and the nonlinear couplings between fluctuations are manifestly shown in $v_n$ and $v_\pi$ (noticing that $\hat{p}$ may contain nonlinear couplings in terms of these basis fluctuations, see also \cite{Gene:1980dk}).

Next, recalling the given definition of the thermodynamic force,
\begin{align}
F_k(\hat{A})=\frac{\delta}{\delta \hat{A}_k}\ln W(\hat{A})
\end{align}
we can translate $\delta (\mu/T)$ and $\delta v$  into $\frac{\delta \ln W(\hat{A})}{\delta \hat{n}}$ and $T_0\frac{\delta \ln W(\hat{A})}{\delta \hat{\pi}}$ according to Boltzmann relation $S\equiv \ln W(\hat{A})$ with $S$ being the total entropy. 

Comparing the relativistic hydrodynamic equations (\ref{Tcon}) to (\ref{heat}) with the Langevin equation,   the kinetic coefficients can also be obtained without effort
\begin{align}
\label{gaa1}
\gamma_{nn}&=-\frac{\sigma T_0(e_0+P_0)}{n_0}  \nabla^2,\\
\label{gaa2}
\gamma_{\pi_i\pi_j}&=-\big(\,(\zeta+\frac{1}{3}\eta)\nabla_i\nabla_j  +\eta\delta_{ij}\nabla^{2}\big),
\end{align}
which demonstrates that
\begin{align}
\Tr(\hat{R}_n(t,\bm{x})\hat{R}_n(t^\prime,\bm{x}^\prime)\,)=-\frac{2\sigma T_0(e_0+P_0)}{n_0} \nabla^2\delta(t-t^\prime)\delta(\bm{x}-\bm{x}^\prime),\\
\Tr(\hat{R}_{\pi_i}(t,\bm{x})\hat{R}_{\pi_j}(t^\prime,\bm{x}^\prime)\,)=-2\big((\zeta+\frac{1}{3}\eta)\nabla_i\nabla_j  +\eta\delta_{ij}\nabla^{2}\big)\delta(t-t^\prime)\delta(\bm{x}-\bm{x}^\prime),
\end{align}
where the spatial coordinate $\bm{x}$ dependence is explicitly recovered. Note throughout the script,   $\bm{x}$ dependence  is frequently suppressed for compactness.

The parametrization given in Eqs.(\ref{gaa1}) and (\ref{gaa2}) is expected to  reproduce the Ornstein–Zernike result \cite{Gene:1980dk,Minami:2011un}.  It is important to note the vanishing of the noise term $R_e$
and its corresponding diffusion kernel. This serves as a nontrivial consistency cross-check. Its vanishing is due to its reversible current being a local density of the conserved quantity, $\delta \hat{T}^{0i}$, which is also included in the list of the chosen relevant slow variables,
\begin{align}
\hat{R}_e\equiv (1-P)iL\hat{e}=(1-P)\partial_i\hat{\pi}_i=\partial_i (1-P)\hat{\pi}_i=0,
\end{align}
where $P\hat{A}=\hat{A}$ is proven in Appendix.\ref{projection}. 

The conclusion that the energy density does not dissipate depends on the choice of the definition of the fluid velocity, or the choice of the set of relevant slow variables in the language of the projection operator . If working in the Eckart frame,   the charge density, rather than the energy density, does not dissipate. Ignoring noise, \cite{Minami:2012hs} gives an illuminating discussion on the frame choice : the Landau frame seems to be a natural choice and  is consistent with the projection operator method. Even for the Eckart frame, the slow dynamics is actually described by the dynamic variables for the Landau frame.

After all is done, we end up with
\begin{eqnarray}
\label{deltan}
\frac{\partial \delta \hat{n}}{\partial t}=&&-\nabla\cdot(\hat{n}\delta \hat{v}) 
-\frac{\sigma T_0(e_0+P_0)}{n_0}  \nabla^2 
\frac{\delta \ln W(\hat{A})}{\delta \hat{n}}+\hat{R}_n(t,\bm{x}),\\
{\tiny} \label{deltae}
\frac{\partial \delta \hat{e}}{\partial t}=&& -\nabla\cdot\hat{\pi} ,\\
\label{pi}
\frac{\partial \hat{\pi}_i}{\partial t}=&& -\nabla_i\delta \hat{p}
-(\zeta+\frac{1}{3}\eta)\nabla_i(\nabla\cdot\frac{\delta \ln W(\hat{A})}{\delta \hat{\pi}_i} ) -\eta\nabla^{2}\frac{\delta \ln W(\hat{A})}{\delta \hat{\pi}_i}+\hat{R}_{\pi_i}(t,\bm{x}) .
\end{eqnarray}


Our results closely resemble those derived in \cite{Minami:2011un} in a classical manner, with the exception of the streaming term in Eq.(\ref{pi}), where the author neglects $\delta p$ and uses a potential condition to rewrite the streaming term. Besides, $\nabla\cdot \hat{\pi}$ replaces $\nabla\cdot( (\hat{e}+\hat{p})\delta \hat{v})$ therein. In this script, we prefer $\nabla\cdot \hat{\pi}$ because the linearity in the basis fluctuations  is unambiguously revealed without causing  confusion.

\end{appendix}

\bibliographystyle{apsrev}
\bibliography{total}

\end{document}